\documentstyle[12pt,aasms4]{article}

\def\gappr{\mathrel{\vcenter{\offinterlineskip \hbox{$>$}
    \kern 0.3ex \hbox{$\sim$}}}}
\def\lappr{\mathrel{\vcenter{\offinterlineskip \hbox{$<$}
    \kern 0.3ex \hbox{$\sim$}}}}

\def\fun#1#2{\lower3.6pt\vbox{\baselineskip0pt\lineskip.9pt
        \ialign{$\mathsurround=0pt#1\hfill##\hfil$\crcr#2\crcr\sim\crcr}}}

\def\ben{\begin{equation}}
\def\be#1{\begin{equation}\label{eq:#1}}
\def\ee{\end{equation}}

\received{}
\revised{}
\singlespace

\begin{document}

\title{
The effect of highly structured cosmic magnetic fields on 
ultra-high energy cosmic ray propagation
}

\author{Gustavo A. Medina-Tanco$^{1,2}$}

\affil{ 1.
Instituto Astron\^omico e Geof\'{\i}sico, University of S\~ao Paulo, Brasil \\
gustavo@iagusp.usp.br
}

\affil{ 2. Dept. of Physics and Astronomy, University of Leeds, 
Leeds LS2 9JT, UK}

\singlespace

\begin{abstract} 

The possibility that the magnetic field is strongly correlated with the 
large-scale structure of the universe has been recently considered in the 
literature. In this scenario the intergalactic magnetic field has a strong 
($\mu$G) regular component spanning tens of Mpc but localized in sheets 
and filaments, while the vast voids in between are almost free of magnetic 
field. If true, this could have important consequences on the propagation 
of ultra-high energy cosmic rays, and severely affect our capacity of doing 
astronomy with charged particles. A quantitative discussion of these effects
is given in the present work.


\end{abstract}

\clearpage
                              
\section{Introduction}

Recently, Ryu, Kang and Biermann (1998) proposed a large-scale scenario in 
which cosmic magnetic fields are clumped inside cosmological sheets and 
filaments in much the same way as the luminous matter does. Furthermore, 
their simulations suggest that the IGMF presents also a remarkable degree 
of alignment along cosmological structures for several Mpc. For such a 
model to be able to account for the existing rotational measure of 
extragalactic sources (e.g., Kronberg 1994), the intensity of the IGMF 
must be relatively high inside cosmological structures ($\sim 1 \mu$G) and 
essentially negligible in the vast voids surrounding those structures. 

This is very different from the common view in which the IGMF is roughly 
uniform in intensity, scaling smoothly with luminous matter density (e.g., 
$B_{IGM} \propto n^{0.2-0.3}$ - Vall\'ee 1997, Medina Tanco 1997b), 
while the reversal 
scale is varied accordingly to account for the observed rotational measure. 
In the latter scenario, the resulting space acquires a cell-like structure 
(e.g., Medina Tanco, Gouveia Dal Pino and Horvath 1997b). The IGMF is more 
evenly distributed and its intensity ranges from few 
$\times 10^{-10}$ G to few $\times 10^{-9}$ G. UHECR particles random walk 
from their sources to the detector through this IGMF configuration producing 
relative small intrinsic angular error boxes (e.g., 
Medina Tanco, Gouveia Dal Pino and Horvath 1997b;
Medina Tanco 1997b,c;
Medina Tanco 1998). With high enough statistics, there is then hope for a 
future UHECR-astronomy. 

In fact, large earthbound projects, like HiRes and Auger, and the OWL twin
satellites are being proposed having such UHECR astronomy among their main 
priorities. However, an IGMF structure as proposed by Ryu and collaborators 
can, if correct, have profound consequences in our understanding of the 
problem of UHECR propagation in the intergalactic medium and on the kind 
of science one can expect to achieve with the new generations of detectors.

This work addresses the problem in a simplified, yet illustrative way. The 
UHECR emission coming from a flat cosmological structure, representative of 
a number of walls that can be encountered in the local universe (see, for 
example, Fairall 1998 and references there in), is analyzed. The dependence 
of flux, angular 
deflection, traversed path and other relevant parameters is studied as a 
function of the relative orientation between the observer, the cosmological 
structure where the sources are embedded and the IGMF. The effects on the 
observed chemical composition are also briefly discussed. 

\section{Numerical calculations and discussion of results}

Ryu and co-workers model for the IGMF is remarkable from the UHECR point 
of view for at least three characteristics: (a) the high intensity of the 
IGMF inside walls and filaments, (b) its laminar structure and large-scale 
order and (c) the marked intensity contrast between the IGMF inside and 
outside walls separated by a relatively thin interface.

In order to analyze the effect of a large scale ordering of the 
intergalactic magnetic field (IGMF) on the observed sky in UHECR, an 
isolated wall immersed in a surrounding void is simulated. In this 
simplified scheme the wall is represented as a disk-like slab of 5 Mpc 
thickness and 20 Mpc radius. The slab is located in the x-z plane and 
centered in the coordinate's origin (see figure 1). 

The space is permeated by a magnetic field with regular component oriented 
along the z-axis. The magnitude of the regular component inside the wall 
is constant and equal to $B_{wall}$, decaying to a very low value, 
$B_{void}$, inside the void. The transition between $B_{wall}$ and 
$B_{void}$ is exponential and takes place over a distance $\Delta y$ 
measured perpendicularly to the wall. In order to break the unrealistic 
homogeneity of the field, a random component, $B_{RND}$, superimposed on 
the regular component, $B_{reg}$, is assumed. $B_{RND}$ is characterized 
by an amplitude $\eta = (B_{RND} / B_{reg}$), power law spectrum 
$\propto \kappa^{-\xi}$, with $\xi = 5/3$, minimum wavenumber 
$k_{min}= 2\pi / 5$ Mpc$^{-1}$ and maximum wavenumber 
$k_{max}= 2\pi / 100$ kpc$^{-1}$.

UHECR particles are injected isotropically with power law spectrum in 
total energy, $dN/dE \propto E^{-2}$, at 100 sources randomly distributed 
inside the volume of the wall. The particle sources have all the same 
integrated luminosity. No sources are considered in the surrounding void.

In what follows only protons are taken into account and energy losses due 
to redshift, pair production and photo-pion production in interactions with 
the cosmic microwave background radiation field are included as given by 
Berezinsky and Grigor'eva (1988).

In figures 2 to 4 the results for $B_{wall} = 0.1 \mu$G, 
$B_{void} = 10^{-10}$ G, $\eta = 0.3$ and $\Delta y = 5$ Mpc are shown. 
Note that, the value chosen for the magnetic field inside the wall is an 
order of magnitude below the values actually estimated by Ryu, Kang and 
Biermann (1998). After normalizing their simulations by the present limit 
in rotation measure (Kronberg 1994), they obtain $\sim 1 \mu$G on 
large-scales outside galaxy clusters.

For illustrative purpose,
figure 2 shows the projections of the orbits of some individual particles 
onto the x-y plane. The general direction of the magnetic field (the regular 
component) is oriented perpendicular to the plane of the figure. All the 
particles were injected with $E=10^{20}$ eV at the same source in the origin
of coordinates, forming an angle of $5^{o}$ with the x-y plane. The only 
thing that varies from run to run is the azimuthal angle at injection, 
measured on the x-y plane with respect to, say the y-axis. It can be seen 
that the dynamics of the UHECR is completely different from a random walk 
and that a large-scale pattern of flow is to be expected. First of all, we 
see that particles tend to drift along the surface of the wall as soon as 
they reach the interface where the magnetic field decreases to its lower 
void value. This combines with a very efficient transport of the particle's 
guiding centers along the z-axis, to give a pattern of UHECR flow that is 
mostly confined to the wall and preferentially directed along the general 
direction of the magnetic field. Furthermore, those particles that escape 
from the surface of the wall tend to do so in a particular direction that 
depends on the side of the wall that is being considered.  Particles 
escaping towards $y > 0$ tend to fly in the $x > 0$ direction, while 
the opposite occurs in the $y < 0$ side of the wall. 
The particular injection point chosen for this example is near enough
to the border of the wall to allow the escape of most of the cosmic rays 
produced there out to the surrounding void. Nevertheless, this escape
is not isotropic.
Furthermore, particles
originated in sources located further inside the wall, will be mainly 
constrained to travel along the IGMF.
This implies that 
a strong anisotropy must be expected in the UHECR emission originated 
from inside the wall. Consequently, the effects of a large-scale regular 
component of the IGMF in the observational properties of UHECR should be 
very dependent on the relative orientation between the ordered magnetic 
field inside the structure and the position of the detector.

The latter is clearly illustrated in figure 3, where several average 
quantities are shown for UHECR impinging a spherical surface of $20$ Mpc 
radius centered in the origin of coordinates. This spherical boundary 
completely encloses the wall, intersecting its borders at the equator 
(see the sketch in figure 1). A spherical coordinate system is chosen 
in which the y-axis defines the polar axis and the z-axis is the origin 
for the longitudes, $\phi$, measured on the plane of the wall (z-x). 
Latitudes, $\theta$, are measured from the plane of the wall. The sphere 
is projected onto the plane of the page using an Aitoff equal area 
projection (Greisen 1993) in which the wall is the horizontal central 
band. The center of the plot is the point where the z-axis intersects 
the sphere ($\phi = 0^{o}$, $\theta = 0^{o}$) and so, in that region, 
the magnetic field emerges perpendicularly to the plane of the figure. 
The UHECR were isotropically injected with a power law energy spectrum 
($dN/dE \propto E^{-2}$ and $E > 4 \times 10^{19}$ eV) at $100$ sources 
randomly located inside the wall and, as previously mentioned, all the 
relevant energy losses were taken into account during propagation.

Figure 3.a shows the logarithm of the flux of UHECR  (in arbitrary 
units). It can be seen that the charged particle flux emerging from 
the wall is very anisotropic, varying in almost three orders of magnitude 
over $4\pi$ sr. The flow of particles in the system, however, shows a 
definite pattern. Most of the UHECR flow along the regular component of 
the IGMF, producing the maxima observed near 
($\phi = 0^{o}$, $\theta = 0^{o}$) and 
($\phi = \pm 180^{o}$,$\theta = 0^{o}$). The relatively few protons 
ejected into the surrounding void, fly away in opposite directions 
above and below the plane of the wall, producing an uneven (and 
antisymmetric) coverage in each hemisphere. In particular, the 
emergent flux is a minimum when observing along the polar axis, 
perpendicular to the wall (i.e., from $\theta = 90^{o}$).  
The drift perpendicular to the IGMF 
direction combines with the flow along the IGMF to produce minima at 
($\phi = \pm 90^{o}$, $\theta = 0^{o}$), i.e., at the border of the 
slab but perpendicular to the magnetic field. Therefore, the emission 
of UHECR from an active large-scale structure can be either considerably 
amplified or severely hampered depending on the relative position of the 
observer. Broadly speaking, a boost is obtained along the IGMF and a drop
in intensity in the direction perpendicular to the IGMF; a wall can be 
invisible for an observer located along a perpendicular through the 
center of the wall.

Figure 3.b shows the deflection between the arrival direction of a typical 
particle and the direction to the source where it originated. It can be 
seen that the deflection angles are very large and so any directional 
information is lost, except for observers looking at the wall flat on. 
Nevertheless, even in the latter situation, the expected error boxes are 
$10-20^{o}$ for most orientations. This could seriously hinder any 
possibility of astronomy with UHECR.

Figure 3.c shows the average deflection between the injection direction 
of a UHECR at the source and its arrival direction at the detector. 
Again, the expected angles are very large, around 90° for most of the 
sky, indicating that any such information is lost. Therefore, if 
acceleration mechanisms that require beaming are involved, it would 
be hopeless to look for some correlation between $\gamma$ and UHECR 
counterparts of a given event 
(e.g., a Gamma Ray Burst - c.f. Stanev, Schaefer and Watson 1996). 
This can represent a test to the 
hypothesis of high intensity large-scale IGMF confined to thin sheets. 
If both, $\gamma$ and UHECR are received in correlation from a single 
source known to have beaming (e.g., a blazar) inside a wall, then the 
IGMF in the wall is most likely highly randomized and well below the 
$0.1 \mu$G scale.

In figure 3.d, the path lengths in Mpc traversed by the particles from 
the source to the sphere are displayed. Exception made of two spots 
antisymmetrically located at each side of the wall, the paths of the 
UHECR are much larger than the actual distance to their sources. This 
has important consequences regarding UHECR composition: if the primaries 
are heavy nuclei, they will more likely arrive as protons regardless the 
position of the detector, even for the new lower estimates of the cosmic 
infrared background (Stecker 1998, Malkan and Stecker 1998; see, however,
Epele and Roulet 1998).

Figures 4.a-c show the Aitoff projections of the sky in UHECR originated 
inside the wall  (roughly a hemisphere), as reconstructed by observers 
located in points a, b and c of figure 1 respectively. Logarithmic fluxes 
are represented in the contour plots and different normalizations are used 
for each one of the three figures. If the observer is located at the border 
of the wall and the large-scale IGMF is aligned with the zenithal direction 
(figure 4.a), the wall is visible as a huge twisted structure covering most 
of the sky. The rotation of the image in UHECR with respect to the wall is 
an indicative of the sense of the IGMF: pointing towards the observer if 
counter-clockwise, pointing away from the observer in the opposite case. 
On the other hand, the image of the wall is completely different when the 
zenith of the observer (located at the border of the wall and in its 
central plane), runs perpendicular to the IGMF (figure 4.b). In this 
case there is a deep minimum right at the zenith bounded by two parallel 
laminar structures that run roughly along the wall. 
Figure 4.a or 4.b (depending on the relative orientation of the IGMF)
should be representative of the Centaurus wall. This feature runs
edge on along the supergalactic plane (de Vaucouleurs 1956)
 dominating the Southern Hemisphere
sky, and will be in full view of the $10^{3}$ km$^{2}$ Auger observatory.
Figure 4.c corresponds 
to an observer outside the wall and watching at it flat on. The image in 
UHECR is in this case a pair of blobs filling most of the sky, and separated 
by a pronounced minimum that cuts the wall in the middle perpendicularly to 
the IGMF direction. Furthermore, the resolution of the sky maps 4.a-c is 
approximately $6^{o}$ and, for that resolution, no individual sources are 
detected.

It should be noted that the IGMF intensity assumed in the calculations 
presented above is one order of magnitude below the upper limit quoted 
by Ryu, Kang and Bierman (1998). The present analysis is therefore a 
conservative one in that context. 
However, 
the implications of such IGMF topology for the analysis of both existing 
and proposed UHECR experiments data must be seriously considered.

\section{Conclusions}

The quantitative results presented heretofore are model dependent and 
so are only intended as illustrative. 
Nevertheless, it has been shown that a relatively well ordered IGMF 
($B_{RND}/B_{reg} \sim 0.3$) compressed to high intensities ($\sim 0.1 \mu$G) 
inside cosmological walls and almost negligible inside voids, dramatically 
changes the picture of UHECR propagation with respect to what has been 
normally assumed in the literature so far. In particular:

\begin{itemize}

\item[1] The UHECR flux leaving a wall can vary by 2 or 3 orders of magnitude 
over $4\pi$, making any conclusion very dependent on the relative orientation 
of the flattened structure containing the sources, the IGMF inside it and the 
observer.

\item[2] Except for privileged observers located near the central 
perpendicular to the wall, directional information is mostly erased, 
hampering source identification.

\item[3] Any information regarding the injection direction of the UHER is 
lost, which means that is should be impossible to detect $\gamma$ and 
UHECR counterparts of a same bursting event if beaming is involved in 
the production of the primaries.

\item[4] The paths traversed by the particles in their way to the detector 
are so large that probably no heavy primary would survive, even for the 
new lower estimates of the infrared background (Malkan and Stecker 1998, 
Stecker 1998).

\item[5] The incoming flux produces a rather flat, featureless contribution 
at the observer's sky. In fact, at a $\sin 5^{o}$ resolution, a density of 
point-like sources as low as $\sim 10^{-2}$ Mpc$^{-3}$, gives no signal 
that could be identified as an individual source on the observed sky 
facing the wall. 

\item[6] The best place to test the correctness of this kind of IGMF model 
is the Southern Hemisphere. There, the huge Centaurus wall (home to the 
Great Attractor and Abell 3627) dominates the sky running roughly coincident 
with the Supergalactic plane and extending out to probably 6000 km/sec 
($\sim 90$ Mpc for $h=0.65$). An additional observational bonus of the 
Centaurus wall is the fact that it not only seen edge on but also 
bordered by the very large local void. Fortunately, this large-scale 
structure will be in the field of view of the $10^{3}$ km$^{2}$ southern 
site of the Auger experiment.

\end{itemize}


This work was done with the partial support of the Brazilian agency FAPESP.

\newpage

\noindent
{\bf Figure Captions}

\bigskip

{\bf Figure 1:} Schematic representation of the model used in the simulations. 
The horizontal slab is the wall containing randomly located UHECR point 
sources. It is there that the IGMF reaches its highest intensity 
$\sim 0.1 \mu$G. The IGMF has two components. The regular component 
is parallel to the z-axis everywhere, its intensity is high and constant 
inside the wall and decays exponentially through an interface to a very 
low value inside the surrounding void. The random component has a power 
law spectrum and scales everywhere as $\eta = B_{RND}/B_{reg} =$ const. 
Protons are injected at the sources inside the slab with a power law 
($dN/dE \propto E^{-2}$ for $E > 4 \times 10^{19}$ eV) spectrum and 
propagated through the system until they reach the spherical border. 
This surface is the site of the different observers considered in the 
work. Several magnitude measured over this surface will be Aitoff 
projected in figures 3a-d, while figures 4a-c show the hemisphere of 
the sky facing the wall as seen by observers located in points a, b and 
c respectively.


{\bf Figure 2:} Trajectories of several test protons injected at $10^{20}$ eV 
at origin of coordinates projected onto the x-y plane. The regular 
component of the magnetic field is perpendicular to the plane of the 
figure. All the protons were injected at an angle of $5^{o}$ with respect 
to the x-y plane, and differ only in the azimuthal injection angle. See 
the text for further details.


{\bf Figure 3:} Aitoff projections of: (a) the logarithm of the UHECR flux 
(arbitrary units), (b) deflection angle (deg) between arrival direction 
and true source position, (c) angle between the arrival direction and 
the injection direction (deg) and (d) path traversed by the UHECR in Mpc 
as measured over the spherical border enclosing the system (see figure 1).


{\bf Figure 4:} Aitoff projection of the hemisphere of the sky facing the wall 
as seen by observers located in positions labeled as a, b and c in 
figure 1. The quantity plotted is the logarithm of UHECR flux and the 
normalization used is different for each case.



\end{document}